\documentclass[10pt,aps,prd,amsmath,floats,floatfix, twocolumn, superscriptaddress,nofootinbib,showpacs,longbibliography]{revtex4-1}
\usepackage[T1]{fontenc}
\usepackage[utf8]{inputenc}
\usepackage{lmodern}
\usepackage{mathtools}
\usepackage{enumitem}
\usepackage{verbatim}
\usepackage{multirow}
\usepackage[dvipsnames, usenames]{xcolor}
\definecolor{linkcolor}{rgb}{0.0,0.3,0.5}
\usepackage[hypertexnames=false, unicode, colorlinks=true, linkcolor=linkcolor,
citecolor=linkcolor, filecolor=linkcolor,urlcolor=linkcolor,
pdfusetitle]{hyperref}
\usepackage{orcidlink}
\usepackage{natbib}
\usepackage[all]{hypcap}
\usepackage{graphicx}
\usepackage{amssymb}
\usepackage[normalem]{ulem} 
\usepackage{bm} 
\usepackage[english]{babel}
\usepackage{blindtext}
\usepackage{ragged2e}
\usepackage[labelfont={small,bf}, font=footnotesize, justification=justified,format=plain,singlelinecheck=off]{caption}

\usepackage{csquotes}
\usepackage{float}

\usepackage[font=footnotesize]{subcaption}
\usepackage{amsmath}
\usepackage{physics}
\usepackage{xspace}
\usepackage[]{mdframed}
\def\6{\partial}
\begin{document}

\newcommand\myeq{\stackrel{\mathclap{\tiny\mbox{poles}}}{=}}

\newcommand{\rg}[1]{\textcolor{red}{\it{\textbf{Rajes: #1}}}}
\newcommand{\cs}[1]{\textcolor{blue}{\it{\textbf{Chiranjeeb: #1}}}}


\title{Can Rotating Black Holes Have Short Hairs?}

\author{Rajes Ghosh}
\email{rajes.ghosh@icts.res.in}
\affiliation{International Centre for Theoretical Sciences, Tata Institute of Fundamental Research, Bangalore 560089, India}

\author{Chiranjeeb Singha}
\email{chiranjeeb.singha@iucaa.in}
\affiliation{
Inter-University Centre for Astronomy and Astrophysics, Pune 411007, India}
 
\date{\today}

\begin{abstract} 
Despite the no-hair theorem, several notable hairy black hole (BH) solutions exist in both General Relativity and modified gravity theories. For such hairs to be detectable, they must extend sufficiently beyond the event horizon. This idea has been rigorously formalized by the no-short-hair theorem, which dictates that all existing hairs of a static spherically symmetric BH must extend at least to the innermost light ring (LR). However, the theorem's applicability to the astrophysically relevant rotating BHs remains elusive as yet. To address this gap, we examine its validity for rotating BHs in the Konoplya-Rezzolla-Zhidenko-Stuchlík and Johannsen classes. Interestingly, for Klein-Gordon separable BHs in these classes that are solutions of non-vacuum GR, the no-short-hair property continues to hold. However, unlike in static cases, this result may not apply in a theory-agnostic fashion due to the rotation-induced repulsive effects. Consequently, we identify a minimal set of additional criteria on the metric and matter content needed for such a generalization in other theories. Our study marks an important first step toward establishing general results on the extent of rotating BH hairs, reinforcing their observational detections. Further extension of this novel result for rotating horizonless objects is also discussed.

\end{abstract}

\maketitle

\section{Introduction}
Contrary to Bekenstein's celebrated no-hair theorem stating that both static and rotating black holes (BHs) cannot be endowed with exterior scalar-meson, massive vector-meson, and massive spin-2 meson fields~\cite{Ruffini:1971bza, 2010GReGr..42..653C, Bekenstein:1971hc, Bekenstein:1972ky}, several hairy BH solutions are well known in General Relativity (GR)~\cite{Bizon:1990sr, Luckock:1986tr, Kanti:1995vq, Campbell:1991rz}. Characterizing such hairy BH spacetimes requires parameters beyond mass, electric charge, and angular momentum. These BHs have been extensively studied in the literature for their unique observational features. Recently, with advances in astrophysical observations, the study of hairy BHs has expanded beyond GR to modified gravity theories, including dilatonic and scalar-Gauss-Bonnet gravity~\cite{Kanti:1995vq, Sotiriou:2013qea, Sotiriou:2014pfa, Julie:2019sab, East:2021bqk, Capuano:2023yyh}.

Despite the existence of hairy BH solutions, detecting these hairs poses a unique challenge. For any hair to be observable, it must not be short/microscopic staying confined only in the near-horizon region without affecting measurable quantities far from the BH. This idea was first formalized in terms of the \textit{no-short-hair theorem} within GR~\cite{Nunez:1996xv, Hod:2011aa}. It asserts that any hair on a four-dimensional, static, spherically symmetric, asymptotically flat BH in GR must extend at least to the innermost light ring (LR), given certain conditions on matter fields. This theorem has since been generalized in a theory-agnostic and dimension-independent manner~\cite{Ghosh:2023kge} (see Ref.~\cite{Tripathy:2024ajw} for an illustration in Lovelock gravity).

These results have significant observational implications. If BH hairs were confined only to near-horizon regions, they might evade detection, misleadingly supporting the no-hair theorem. However, the no-short-hair theorem provides an optimistic lower bound, indicating that probing up to the innermost LR is enough to detect BH hairs. If no hair is found within this region, particularly via lensing or shadow observations, the presence of hair can be confidently ruled out.

A critical gap remains in the no-short-hair theorem, as its applicability to rotating astrophysical BHs remains elusive. Rotation complicates the situation through a “centrifugal repulsive effect,” which disrupts the balance in matter configurations at asymptotic regions (where hairy fields would radiate away) and near the horizon (where fields would have been absorbed). This repulsion could potentially support short hairs on rotating BHs, depending on the theory and matter involved~\cite{Hod:2014sha}. While these competing effects are not yet fully understood, their theoretical and observational importance is undeniable. Inspired by these considerations and previous work on scalar hairs~\cite{Hod:2016ixl}, we aim to investigate the no-short-hair theorem for rotating BHs.

To this end, we examine two well-studied phenomenological models of rotating BHs: the Konoplya-Rezzolla-Zhidenko-Stuchlík (KRZS)\cite{Konoplya:2016jvv, Konoplya:2018arm} and Johannsen metrics~\cite{Johannsen:2013szh}. These models offer a unified, theory-agnostic framework for exploring rotating BHs. Given that the no-short-hair theorem’s universality might be compromised~\cite{Hod:2014sha}, we aim to identify a minimal set of criteria on both the metric and matter content to generalize this theorem. One key element, already built into the KRZS class, is the separability of geodesics and the Klein-Gordon (KG) equation, crucial for simplifying the analysis and isolating field behavior in BH spacetime. We illustrate our findings for rotating BHs in GR and discuss possible extensions for rotating horizonless objects.

An important distinction between rotating and non-rotating cases is that the extension of the BH hair is no longer connected to the innermost LR of rotating BHs. This departure suggests that the ``no-shortness'' of rotating BH hairs may not hold unconditionally. Specifically, we demonstrate that for the Johannsen class without KG separability, the no-short-hair theorem may not apply. This implies that the extent of rotating BH hairs depends on more complex conditions beyond simply reaching the innermost LR. Accordingly, we establish the following theory-agnostic result.  
\begin{mdframed}
\textit{Theorem.} All existing hairs of a rotating KRZS BH (equivalently, KG separable Johannsen BH) must extend outside $r_*=\text{min}\{r_0,r_0'\}$ at least at the poles, provided the matter ($T^\mu_\nu$) obeys WEC, non-positive trace condition, and the energy density falls faster than $r^{-4}$ at large radial distances. Here, $r_0$ characterizes the LR of an auxiliary spacetime constructed in a subsequent section and $r_0'$ represents the radius till which $\partial_\theta \left(\sqrt{-g}\, T^\theta_r\right) \geq 0$ at poles.
\end{mdframed}

In other words, if the stress-energy tensor obeys the aforementioned conditions, all hairs of a rotating KRZS BH or a KG separable Johannsen BH will be bigger than a certain radius $r_*$ at the poles. In the special case of $\partial_\theta \left(\sqrt{-g}\, T^\theta_r\right) = 0$, as in GR, the minimum hair extent is $r_*=r_0$, upholding the no-short-hair property for GR BHs in the KRZS class. We focus on the poles, where the rotation axis intersects the horizon, to minimize rotation-induced repulsion. The full angular profile of the hair around the rotating BH warrants further investigation. Nonetheless, our study marks a key first step toward generalizing the no-short-hair theorem for rotating BHs.

\section{no-short-hair Theorem for Static Hairy Black Holes: A review}
Before discussing the possible extension of the no-short-hair theorem for rotating BHs, let us begin by reviewing a few important results of Refs.~\cite{Nunez:1996xv, Hod:2011aa, Ghosh:2023kge} for the static spherically symmetric case. This will serve as a warm-up and make our analysis self-contained. Historically, this theorem was first established assuming GR field equations and $4$-spacetime dimensions~\cite{Nunez:1996xv}. The proof starts by writing a static spherically symmetric metric as
\begin{equation}
    ds^2 = -e^{-2\delta(r)}\mu(r)dt^2 + \frac{dr^2}{\mu(r)} + r^2d\Omega_{(2)}^2 \ , \nonumber
\end{equation} 
representing a hairy BH solution of GR with a stress-energy tensor $T^\mu_\nu$, chosen in such a way that can indeed support hairs. In particular, one assumes that $T^\mu_\nu$ obeys WEC, non-positive trace condition, and the energy density falls faster than $r^{-4}$ at large radial distances~\cite{Nunez:1996xv}. 

Then, what is the minimal radius of these existing hairs, i.e., till which extent the radial pressure $p_r=T^r_r$ must be non-zero? To derive an answer, one uses Einstein's field equations and writes the radial component of the conservation equation $\nabla_\mu T^\mu_r = 0$ in the following suggestive manner:
\begin{equation}
    (r^4 p_r)' = \frac{r^3}{2\mu} (3\mu - 1 - 8\pi r^2 p)(\rho+p) + T\,r^3 \ , \nonumber
\end{equation}
where $(^\prime)$ represents radial derivative, $\rho = -T^t_t$ is the energy density, and $T$ denotes the trace of the stress-energy tensor.

Now, following Ref.~\cite{Nunez:1996xv}, one analyzes the behavior of $(r^4 p_r)$ in the near-horizon ($r \sim r_H$) region and imposes regularity to deduce
$p_r|_{r_H} \leq 0$, and $p_r'|_{r_H} < 0$. These inequalities along with asymptotic flatness and aforementioned energy conditions, in turn, ensure that $(r^4 p_r)$ is a non-positive and monotonically decreasing function at least up to a certain radius away from the horizon. The geometrical significance of this radius was first recognized in Ref.~\cite{Hod:2011aa} as the innermost LR. Consequently, all existing hairs of a non-rotating BH in GR must extend till the innermost LR. In fact, as an implication of the spherical symmetry, one can easily obtain the complete angular profile of the ``hairosphere''.

The above proof relies heavily on the framework of GR and the dimensionality of the spacetime being four. However, later in Ref.~\cite{Ghosh:2023kge}, the above theorem was generalized in a theory-agnostic and dimension-independent manner. As a result, static hairy BHs must obey the no-short-hair theorem independent of the underlying gravitational theory. However, as discussed earlier, a significant gap persists in the applicability of this theorem for rotating BHs, which we aim to bridge in the subsequent sections. Our goal will be to proof the boxed theorem in the Introduction. 

\section{Rotating hairy Black Holes}
Any rotating $4$-dimensional stationary, axisymmetric, and circular BH metric can be expressed in terms of Boyer-Lindquist coordinates $\{t, r, \theta, \varphi\}$ as~\cite{Wald:1984rg},
\begin{equation} \label{gmetric}
    ds^2 = g_{tt}\, dt^2 + g_{rr}\, dr^2+g_{\theta \theta}\, d\theta^2 + 2\, g_{t \varphi}\, dt\, d\varphi + g_{\varphi \varphi}\, d\varphi^2\, ,
\end{equation}
where all metric coefficients are functions of $(r,\theta)$ alone. The determinant of this metric is $g = - \Delta\, g_{rr}\, g_{\theta \theta}$ with $\Delta = g_{t \varphi}^2 - g_{tt}\, g_{\varphi \varphi}$. Assuming the energy-momentum tensor $T_{\mu \nu}$ sourcing this metric respects spacetime symmetries, $T_{\mu \nu}$ depends only on $(r,\theta)$. Moreover, due to the metric’s circular nature, $T_{\mu \nu}$ must be invariant under simultaneous inversion of $(t, \varphi)$ coordinates, allowing only $T_{r \theta}$ and $T_{t \varphi}$ as non-diagonal components. Therefore, the stress-energy tensor has six independent components, namely energy density $\rho:= -T^t_t$, radial pressure $p_r:= T^r_r$, angular pressures $\{p_\theta:= T^\theta_\theta,\, p_\varphi:= T^\varphi_\varphi\}$, and only two cross-components $\{T^\theta_r,\, T^t_\varphi\}$. All other components are either derived from these or are zero.

Now, to derive the central equation for our study, let us consider the radial component of the stress-energy conservation: $\nabla_\mu T^\mu_r = 0$. After some tedious algebraic manipulations, we obtain 
\begin{equation} \label{pr}
    \begin{split}
        &\frac{1}{g_{\theta \theta}^2}\, \partial_r\left(g_{\theta \theta}^2\, p_r\right) = - \frac{1}{\sqrt{-g}}\, \partial_\theta \left(\sqrt{-g}\, T^\theta_r\right)+\frac{1}{2}\left(3\, p_r + p_\theta \right)\, T_1 \\
        & - \left(\rho + p_r \right)\, T_2  +\frac{T_3}{2}\, T - \frac{T^t_\varphi}{2\Delta}\, g_{tt}\, g_{t\varphi}\, \partial_r \log \left(1+\frac{\Delta}{g_{tt}\, g_{\varphi \varphi}}\right)\, ,
    \end{split}
\end{equation}
where we have used the following notation:
\begin{equation} \label{T123}
    \begin{split}
        &T= g^{\mu \nu}\, T_{\mu \nu}=-\rho+p_r+p_\theta+p_\varphi\, ,\, T_1 = \partial_r \log \left(\frac{g_{\theta \theta}}{g_{\varphi \varphi}}\right)\, ,\\ &T_2 = \partial_r \log \left(\frac{\sqrt{\Delta}}{g_{\varphi \varphi}}\right)\, ,\, T_3 = \partial_r \log \left(g_{\varphi \varphi}\right)\, .
    \end{split}
\end{equation}
One can easily verify that Eq.~\eqref{pr} reduces to the familiar static spherically symmetric case~\cite{Ghosh:2023kge}, where all quantities become functions of $r$ alone, $\{T_1, T^\theta_r, T^t_\varphi\}$ vanish, and $T_2$ becomes proportional to the LR equation. Consequently, the no-short-hair theorem applies in a theory-agnostic manner for non-rotating asymptotically flat BHs discussed earlier, given that the matter obeys the WEC, a non-positive trace condition, and $\rho$ decays faster than $r^{-4}$ at $r \to \infty$~\cite{Ghosh:2023kge}.

However, for the rotating case, even with asymptotic flatness and aforementioned stress-energy conditions, it seems that the no-short-hair theorem does not follow in general~\cite{Hod:2014sha} and theory-agnostically. This is primarily because the RHS of Eq.~\eqref{pr} no longer has a specific sign, unlike in the non-rotating case. Moreover, most surprisingly in the rotating case, the term $T_2$ is not proportional to the LR equation. But, given the importance of this theorem, we now take a closer look to extend the no-short-hair result for the rotating case as well by considering some well-studied rotating BH metrics. 

\section{Black Holes in the KRZS class}
With the above motivation, we shall first study the status of the no-short-hair theorem in the KRZS class~\cite{ Konoplya:2018arm}, which represents the most general stationary, axisymmetric, and asymptotically flat rotating BH spacetime leading to separable KG and geodesic equations. This class has the following metric in Boyer-Lindquist coordinates~\cite{ Konoplya:2018arm}:
\begin{equation} \label{KRZmetric}
    \begin{split}
        ds^2 = &-\frac{N^2-W^2\, \sin^2\theta}{K^2}\, dt^2+ r^2\, K^2\, \sin^2\theta\, d\varphi^2\\
        &-2\, r\, W\, \sin^2\theta\, dt d\varphi +\Sigma\left(\frac{B^2}{N^2} \, dr^2 + r^2\, d\theta^2 \right)\, .
\end{split}
\end{equation}
Here, $a$ is the rotation parameter, and separability requires~\cite{ Konoplya:2018arm},
\begin{equation} \label{KRZfunc}
    \begin{split}
        &\Sigma = R_\Sigma(r)+\frac{a^2\, \cos^2\theta}{r^2},\, N^2=R_\Sigma(r)-\frac{R_M(r)}{r}+\frac{a^2}{r^2},\, \\ 
        & B^2 = R^2_B(r),\, \, W= \frac{a\, R_M(r)}{r^2\, \Sigma(r,\theta)},\,   \\ 
        & K^2=\frac{1}{\Sigma}\left[R_\Sigma(r)^2+R_\Sigma(r) \frac{a^2}{r^2}
        +\frac{a^2\, \cos^2\theta\, N^2}{r^2}\right]  +\frac{a\, W}{r}\, .
\end{split}
\end{equation}
The event horizon (which is also the Killing horizon) is given by $N(r) = 0$. Moreover, the asymptotic flatness is assured if $R_\Sigma, R_B,\, \text{and}\, N^2 \to 1$ as $r\to\infty$~\cite{Konoplya:2018arm}.

\subsection{Theory-agnostic considerations}
Even with this metric, the RHS of Eq.~\eqref{pr} does not have a particular sign for different values of $\theta \in [0,\pi]$. This should not surprise us in the absence of spherical symmetry. In fact, as shown in Ref.~\cite{Acharya:2024kvv}, the no-short-hair property may only hold for certain specific values of $\theta$. Moreover, as explained in the Introduction, the subtle balance of the matter configurations at spatial infinity and near the horizon will be distorted due to BH's rotation. Then, drawing from classical mechanics, the poles will experience minimal rotational effects, suggesting that the no-short-hair property may still apply there. Hence, at the poles, we have
\begin{equation} \label{prKRZ}
    \begin{split}
        &P_r'(r) \, \myeq\, - h^2(r) \left[\frac{\partial_\theta \left(\sqrt{-g}\, T^\theta_r\right)}{\sqrt{-g} }\right]_{\theta \to \{0,\pi\}} \\ &\kern 4em+\frac{\left(\rho + p_r \right)\, h(r)\, L(r)}{2\, f(r)} +\frac{h(r)\, h'(r)}{2}\, T_p\, ,
    \end{split}
\end{equation}
where $P_r(r) = h^2(r)\, p_r(r)$, $T_p = T(\theta \to \{0, \pi\})$, the functions $f(r) = r^2\, N^2(r)/h(r)$, $h(r) = r^2\, R_\Sigma(r)+a^2$, and $L(r)$ given by 
\begin{equation} \label{LR}
    L(r) = f(r)\, h'(r) - h(r)\, f'(r)\, 
\end{equation}
will be of crucial importance to us. We must emphasize that $L(r)$ is not the LR equation for the rotating case. However, for the non-rotating ($a=0$) case, $L(r)$ becomes proportional to the corresponding LR equation for all values of $\theta$.

Since the KRZS metric is symmetric about the equatorial plane, both limits $\theta \to \{0,\pi\}$ yield the same result. In deriving the Eq.~\eqref{prKRZ}, we have used $T^t_\phi \to 0$ at the poles, which follows from the regularity of $T^{\mu}_{\nu}$ and the identity $g_{tt}\, T^{t}_{\varphi} = g_{\varphi \varphi}\, T^{\varphi}_{t} - g_{t \varphi}\, (\rho + p_\varphi)$. Notably, Eq.~\eqref{prKRZ} closely resembles its non-rotating counterpart~\cite{Ghosh:2023kge}. In fact, except for $T^\theta_r$, the other two terms in the RHS of Eq.~\eqref{prKRZ} have specific signs for a range in $r$. To see this, note that $h(r \geq r_H) > 0$ with $r_H$ being the non-extremal event horizon. Then, further assuming $h'(r) > 0$ (convexity)~\cite{Ghosh:2023kge}, we find $\rho+p_r=0$ at the horizon so that RHS remains finite at $r_H$ (for an intuitive argument see Ref.~\cite{Dorlis:2023qug}). In turn, this implies $P_r(r_H)<0$. 

Then, with WEC ($\rho+p_r \geq 0$), non-positive trace condition $T \leq 0$, and $p_r(r)$ falls to zero faster than $r^{-4}$, we observe that $P_r(r)$ is a monotonically decreasing function at least to the innermost positive root of $L(r)$, say $r_0$. In fact, for our purpose, these energy conditions are only required at poles. There indeed exists such a root of $L(r)$ as $L(r_H) < 0$ but $L(r \to \infty) > 0$. Thus, $L(r)$ must have an odd number of positive roots outside the horizon, and if $r=r_0$ is the smallest, then $L(r_H \leq r \leq r_0) \leq 0$. In summary, both the second and third terms in the RHS of Eq.~\eqref{prKRZ} are indeed non-positive in $\mathcal{D}:= [r_H,r_0]$.

As noted earlier, the radius $r_0$ does not correspond to the innermost LR in the rotating case, and we lack a general analytic expression for it. So, what does $r_0$ signify? To address this, we consider an auxiliary static spacetime in $\{t,r,\theta,\varphi\}$ coordinates,
\begin{equation} \label{auxmetric}
    ds_{aux}^2=ds_p^2+ h(r)\, \sin^2\theta\, d\varphi^2\, ,
\end{equation}
where $ds_p^2$ represents the KRZS metric given by Eq.~\eqref{KRZmetric} at poles, and $a$ should be interpreted as an extra parameter affecting the areal distance. Our auxiliary spherically symmetric metric fictitiously extends $ds_p^2$ for all values of $\theta$ by adding $h(r)\, \sin^2\theta\, d\varphi^2$ term. The usefulness of this construction is that this new BH spacetime, with horizon $r_H$, has its innermost LR at $r_0$. For $a=0$ (spherically symmetric case), both the KRZS metric and $ds_{aux}^2$ share the same LRs. However, this is not true for non-zero $a$.

Now, we only need to analyze the sign of the $T^\theta_r$ term at poles. This term may take both signs in $\mathcal{D}$ depending on the matter content. As a result, the no-short-hair property fails for rotating BHs unless $\partial_\theta \left(\sqrt{-g}\, T^\theta_r\right) \geq 0$ at poles in a non-empty domain $\mathcal{D}':= [r_H,r_0']$. This is exactly the sought-after minimal additional criteria required to uphold the no-short theorem for rotating BHs in the KRZS class, acting as an extra restriction on the matter content that can produce hairy BHs. If this condition holds, all existing hairs must extend beyond $r_*=\text{min}\{r_0,r_0'\}$ at the poles (at least). This completes our proof of the result outlined in the Introduction. However, unlike the non-rotating case, $r_*$ generally lacks a universal geometrical interpretation and depends on the specific matter content considered.

\subsection{Illustrations in GR}
In practice, to calculate $T^\theta_r$ and check the aforesaid condition, we need to assume a specific theory. To illustrate it more, let us consider a particularly interesting case where the KRZS metric is a solution of non-vacuum GR. Then, using Einstein's equations, one obtains that $T^\theta_r$ vanishes trivially for all $(r,\theta)$. Hence, $r_0' \to \infty$ and the extent ($r_*$) of hair is characterized only by $r_0$. \textit{Consequently, the no-short-hair theorem holds for rotating KRZS BHs in GR, if the other energy conditions are met.}

Let us now consider two specific examples in GR for concreteness:
(i) As the first example, we consider $R_M(r)$ to be free, and $R_\Sigma(r) = R_B(r) = 1$. For this case, $(\rho+p_r)$ is identically zero, $\rho = V(r)/h^2(r)$, and $T_p = - V'(r)/[r\, h(r)]$ with $V(r) = r^2\, R_M'(r)$. Therefore, in order to satisfy all energy conditions, we require $V(r)$ to be a positive non-decreasing function till some large distance away from the horizon, and after that $V(r)$ must decrease to zero faster than $r^{-1}$ as $r \to \infty$. Note that such a behavior for $V(r)$ is impossible for a monotonic $R_M(r)$, for which no-short-hair property may fail.

(ii) In the above example, we could not find an explicit expression for $r_0$ due to the unspecified form of $R_M(r)$. Hence, let us now consider another example of the KRZS metric specified by a free-function $R_B(r)$, $R_\Sigma = 1$, and $R_M = 2M$.
\begin{figure}[ht!]
    \includegraphics[width=0.9\linewidth]{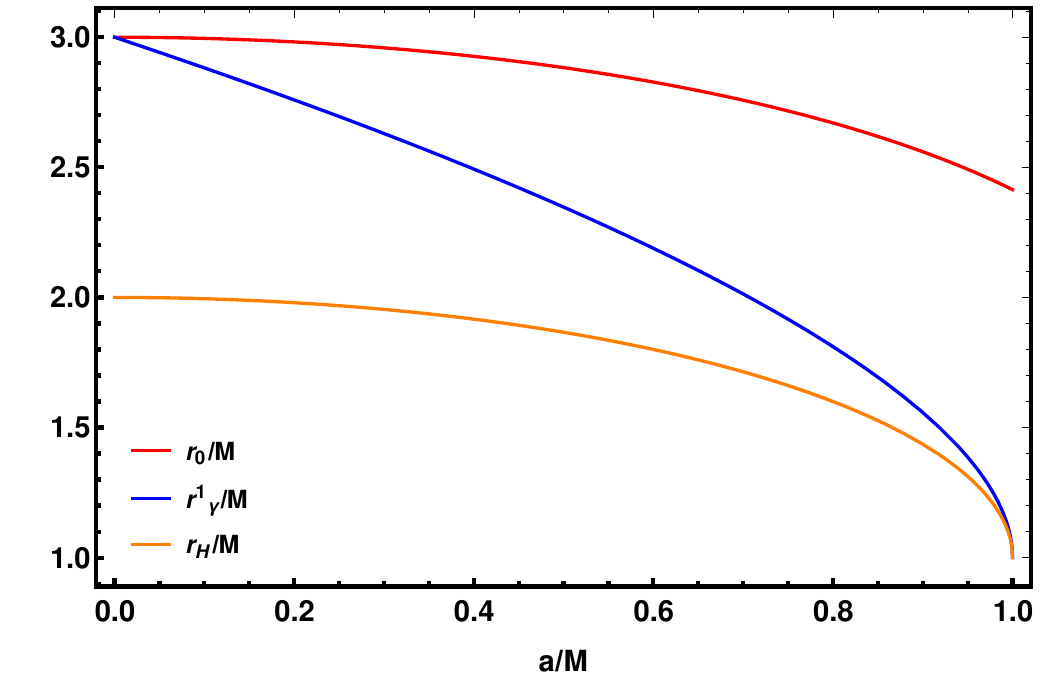}
    \caption{Plots showing radius $r_0$ of existing hairs in red, innermost LR radius $r_\gamma^1$ (but taken at poles) in blue, and horizon radius $r_H$ in saffron. Note that $r_0 \geq r_\gamma^1$ for all $a/M$.} 
    \label{fig:1}
\end{figure}
For this case, we shall avoid writing the explicit expressions for various components of $T^\mu_\nu$, as they are quite big and not particularly illuminating. Now, it turns out that $r_0 = M+A\, \cos(\phi/3)$ is the minimum extent of the hairosphere at poles. Here, $A = 2\, \sqrt{M^2-a^2/3}$, and $\phi = \cos^{-1}[M(M^2-a^2)/(M^2-a^2/3)^{3/2}]$. From Fig.~[\ref{fig:1}], it is clear that the polar extent of hairs is bigger than the innermost LR, provided one can indeed construct a function $R_B(r)$ that satisfies all the required energy conditions. Though there is no a priori reason to believe such functions will not exist, but finding one requires more work that we leave for the future. Note also that although this LR exists in the equatorial plane, we imagine a sphere of its radius centering at the origin to fix a standard scale for comparing hairs' length.

\subsection{BHs in the Johannsen class}
We now study the no-short-hair theorem for rotating BHs in the Johannsen class~\cite{Johannsen:2013szh} assuming GR field equations. It has all the properties of the KRZS class except for the KG separability. However, we shall show this is exactly the condition needed to satisfy the weaker (equality) version of $\partial_\theta \left(\sqrt{-g}\, T^\theta_r\right) \geq 0$ at poles. This condition is sufficient (may not be necessary) to retain the no-short-hair property. A general BH spacetime in the Johannsen class has the following metric components~\cite{Johannsen:2013szh},
\begin{eqnarray} \label{Jmetric}
g_{tt} &=& -\frac{\tilde{\Sigma}[\tilde{\Delta}-a^2A_2(r)^2\sin^2\theta]}{[(r^2+a^2)A_1(r)-a^2A_2(r)\sin^2\theta]^2}, \nonumber \\
g_{t\varphi} &=& -\frac{a[(r^2+a^2)A_1(r)A_2(r)-\tilde{\Delta}]\tilde{\Sigma}\sin^2\theta}{[(r^2+a^2)A_1(r)-a^2A_2(r)\sin^2\theta]^2}, \nonumber \\
g_{rr} &=& \frac{\tilde{\Sigma}}{\tilde{\Delta}\, A_5(r)}\, ,\, \,  g_{\theta \theta} = \tilde{\Sigma}\, , \nonumber \\
g_{\varphi \varphi} &=& \frac{\tilde{\Sigma} \sin^2 \theta \left[(r^2 + a^2)^2 A_1(r)^2 - a^2 \tilde{\Delta} \sin^2 \theta \right]}{[(r^2+a^2)A_1(r)-a^2A_2(r)\sin^2\theta]^2}, 
\end{eqnarray}
where 
$\tilde{\Sigma} = r^2 + a^2 \cos^2\theta + \tilde{f}(r)$, and $\tilde{\Delta} = r^{2}-2 M r+a^{2}$~\cite{Johannsen:2013szh}. Now, at the poles, we have both $T_1=T^t_\varphi\, \myeq \,0$ like in KRZS class, and $T_3 = \tilde{h}'/\tilde{h} > 0$ (convexity) with $\tilde{h}(r) = r^2+a^2+\tilde{f}(r)$, except $\partial_\theta \left(\sqrt{-g}\, T^\theta_r\right) \neq 0$ unlike the KRZS case (no particular sign either). Thus, the no-short-hair theorem does not generally apply to Johannsen BHs. However, let us now consider a subclass with KG separability, requiring~\cite{Konoplya:2018arm}  
\begin{equation}
    \tilde{f}(r) = (r^2+a^2)\left[ \frac{A_1(r)}{A_2(r)} - 1\right]\, . \nonumber
\end{equation}
With this $\tilde{f}(r)$, the above problematic term vanishes and the reduced metric in Eq.~\eqref{Jmetric} can be mapped to the KRZS class as shown in Ref.~\cite{ Konoplya:2018arm}. Then, one can proceed as before for establishing the no-short-hair theorem.

\section{Comments on horizonless compact objects}
So far, we focused our attention only to rotating BHs. Now, we shall try to extend our results for horizonless compact objects as well. Although such horizonless objects can grow hairs, it is not clear what is the angular profile of these hairs and whether they have a minimal length like in the BH case. Thus, examining the conditions needed to hold the no-short-hair theorem is essential. Moreover, several previous works~\cite{Ghosh:2023kge, Peng:2020hkz, Liu:2023swl} have discussed this theorem in the context of static horizonless objects. Hence, it would be interesting to provide a parallel discussion for the rotating case as well.

However, since the KRZS parametrization fails, we shall use the Lewis-Papapetrou metric (in coordinates not to be confused with that of Boyer-Lindquist) given by~\cite{Berti:2015itd, Gervalle:2022fze} 
\begin{equation} \label{LPmetric}
    \begin{split}
        d s^2= &-f\, d t^2+\frac{\ell}{f}\Big[H\left(d r^2+r^2 d \theta^2\right) \\
        &+r^2\, \sin ^2 \theta \left(d \varphi-\frac{w}{r} d t\right)^2\Big]\, , 
\end{split}
\end{equation}
to characterize the spacetime of a horizonless object. Then, geodesic and KG separability require that $\{ f, \ell, w\}$ are functions of $r$ alone and $H=1$~\cite{Papadopoulos:2020kxu, Baines:2023dhq}. Also, the metric must obey other conditions like asymptotic flatness, axial symmetry, equatorial reflection symmetry, and non-existence of conical singularity. They are derived in Sec-(II.B) of Ref.~\cite{Gervalle:2022fze}, which we avoid writing here. 

We shall further assume that this star is a solution of GR with matter obeying the WEC and $\rho$ falls faster than $r^{-4}$ as $r \to \infty$. Then, using the general Eq.~\eqref{pr} for the Lewis-Papapetrou metric and taking the polar limit, we again obtain the same equation as in Eq.~\eqref{prKRZ}, but with $h(r) = g_{\theta \theta}$ and $f(r)$ obtained from Eq.~\eqref{LPmetric}. Moreover, as before, Einstein's equations kill the $T^\theta_r$ term. 

However, in the absence of a horizon, we must put certain conditions as mentioned in Sec-(II.B) of Ref.~\cite{Gervalle:2022fze} on the metric functions at the center ($r=0$) of the star. These conditions perform two purposes, namely they provide the boundary conditions for integrating Einstein's equations, and maintaining regularity of $(\rho, p_r)$ inside the object. Then, we have $L(0) > 0$, and $L(\infty) > 0$ for asymptotic flatness. 

Now, in the spirit of Ref.~\cite{Ghosh:2023kge} and to extend the no-short-hair result for the horizonless case, we need to assume $T \geq 0$, which is exactly the opposite of the BH case. With these conditions, we first notice from Eq.~\eqref{prKRZ} that $P_r'(\infty) > 0$ and hence, $P_r(\infty)<0$ to uphold asymptotic flatness. Therefore, starting with a positive value at $r=0$, $P_r(r)$ must obtain a maximum at some $r_0$, which turns out to be the smallest positive root of $L(r)$. This radius represents the least extent of hairs and has the same physical meaning of the innermost LR of the auxiliary metric in Eq.~\eqref{auxmetric}, which completes our extension of the no-short-hair theorem for rotating horizonless objects.

Let us close this section with a useful remark. Instead of KRZS parametrization, one may use the Lewis-Papapetrou metric for the BH case as well~\cite{Berti:2015itd}. A similar calculation shows that all our previous results about the extent of the hair continue to hold.

\section{Discussion and Conclusions}
In this work, we have thoroughly examined the no-short-hair theorem for rotating BHs in $4$-dimensions. While this theorem holds theory independently for non-rotating cases, our findings indicate that, except in GR, rotating BH solutions may not universally adhere to this result in other theories. Additional criterion are needed for a theory-agnostic generalization. Notably, unlike in static cases, the hairosphere of rotating KRZS BHs extends to a radius that typically differs from the innermost LR. Moreover, we have illustrated the consequence of the aforementioned criterion in the context of GR. In a future work, we plan to understand the physical interpretations of this condition and try to explore a theory-agnostic generalization.

We have also demonstrated that the no-short-hair theorem may not apply to the Johannsen class unless the metric is KG separable, a condition automatically satisfied in static spherically symmetric cases. Since this condition may not hold for arbitrary rotating metrics, further exploration is warranted. Additionally, we investigated the potential extension of the no-short-hair property to horizonless objects. Other possible generalizations could involve relaxing various symmetry assumptions, and altering the asymptotic structure to de Sitter or anti-de Sitter types~\cite{Ishibashi:2023vsm}.

In conclusion, our study represents the first effort to understand the general properties of hairy rotating objects in a unified framework. We hope this work will pave the way for future research into the classification of rotating BHs based on the presence/absence of short hairs. By continuing to investigate these solutions, we aim to deepen our understanding of BHs in astrophysical contexts, facilitating robust detection of hairs and providing new insights into BH physics and gravitational theories.

\section{acknowledgements}
We thank the anonymous referee for their valuable comments on an earlier draft. We also thank Shahar Hod, Daniel Sudarsky, and Sudipta Sarkar for useful comments that have improved both the content and the presentation of the paper.

\end{document}